\documentclass{ws-ijmpe}
 \usepackage{amsmath}
 \usepackage{mathrsfs}
 \usepackage{bm}

\newcommand{\br}{{\mathbf{r}}}
\newcommand{\bJ}{{\mathbf{J}}}

\newcommand{\beq}{\begin{equation}}
\newcommand{\eeq}{\end{equation}}
\newcommand{\beqn}{\begin{eqnarray}}
\newcommand{\eeqn}{\end{eqnarray}}

\begin{document}

\markboth{H. Mei et al.,}{Impurity effect of $\Lambda$ hyperon on
shape-coexistence nucleus $^{44}$S in the energy functional based
collective Hamiltonian.}

\catchline{}{}{}{}{}
%%%%%%%%%%%%%%%%%%%%%%%%%%%%%%%%%%%%%%%%%%%%%%%%%%%%%%%%%%%%%%%%%%%%

\title{Impurity
effect of $\Lambda$ hyperon on shape-coexistence nucleus $^{44}$S in
the energy functional based collective Hamiltonian\footnote{Based on
talk presented at 18th Nuclear Physics Workshop ``Maria and Pierre
Curie", 2011, Kazimierz, Poland} }

\author{\footnotesize H. Mei$^{1}$, Z. P. Li$^{1}$,
J. M. Yao$^{2,1}$\footnote{Corresponding author: jmyao@swu.edu.cn},
K. Hagino$^{3}$}

 \address{$^{1}$ School of Physical Science and Technology, Southwest University, Chongqing 400715 China\\
 $^{2}$ Physique Nucl\'eaire Th\'eorique, Universit\'e Libre de Bruxelles, \\ C.P. 229, B-1050 Bruxelles, Belgium\\
 $^{3}$Department of Physics, Tohoku University, Sendai 980-8578, Japan}

 \maketitle

%********************************************************************%
%                     abstract part
%********************************************************************%
\begin{abstract}

The non-relativistic Skyrme energy density functional (EDF) based
collective Hamiltonian, that takes into account dynamical
correlations related to the restoration of broken symmetries and
fluctuations of quadrupole collective variables, is applied to
quantitatively study the impurity effect of $\Lambda$ hyperon on the
collectivity of $^{44}$S. Several Skyrme forces for both the
nucleon-nucleon ($NN$) and $\Lambda$-nucleon ($\Lambda N$)
interactions are used. The influence of pairing strengths on the
polarization effect of $\Lambda$ hyperon is also examined. It is
found that although these Skyrme forces with different pairing
strengths give somewhat different low-lying spectra for $^{44}$S,
all of them give similar and generally small size of $\Lambda$
reduction effect (within $5\%$) on the collective properties.

\end{abstract}

%********************************************************************%
%                     introduction part
%********************************************************************%
\section{Introduction}
Since the first discovery of $\Lambda$ hypernuclei by observing
cosmic-rays in emulsion chambers,\cite{Danysz53} hypernuclei --
which are nuclei with one or more of the nucleons being replaced
with hyperons -- have been used as a natural laboratory to study
hyperon-nucleon ($YN$) and hyperon-hyperon ($YY$) interactions,
properties of hadrons in nuclear environment, and in particular the
impurity effect of hyperon in nuclear
medium.~\cite{Chrien89,Dover89,Bando90,Hashimoto06} Due to the
absence of Pauli's principle between the nucleon and the $\Lambda$
particle, a $\Lambda$ hyperon can probe deeply into the interior of
nuclear medium and have important influences on its properties,
including softening the equation of state,\cite{Glendenning00}
modifying the shape and size of finite nucleus,~\cite{Tanida01}
changing the nuclear binding and thus the driplines of neutrons and
protons~\cite{Samanta06} as well as the fission barrier heights in
heavy nuclei.~\cite{Minato09} The facilities built at J-PARC will
provide an opportunity to perform hypernuclear $\gamma$-ray
spectroscopy study with high precision by improving the quality of
the secondary mesonic beam.\cite{Tamura09} These facilities offer
useful tools to study the low-lying states of hypernuclei.

In recent years, both the non-relativistic Skyrme-Hartree-Fock (SHF)
~\cite{Zhou07,Schulze10,Win11} and the relativistic mean-field
(RMF)~\cite{Win08,Lu11} approaches have been applied to study the
polarization effect of $\Lambda$ hyperon on the deformation of
atomic nuclei. It has been found that, generally, the shape
polarization effect of the $\Lambda$ hyperon is not evident, but
with several exceptions, including $^{13}_\Lambda$C,
$^{23}_\Lambda$C, and $^{29,31}$$_\Lambda$Si. In these studies,
however, the predicted energy surface is shown to be somewhat soft,
in which case a large shape fluctuation effect of collective
vibration might be expected. Furthermore, symmetry (e.g.,
translation, rotation, particle number) is usually spontaneously
broken in the single-reference (SR) EDFs. The symmetry restoration
becomes particular important for studying the spectrum of low-lying
states.

Recently, the low-lying states of $\Lambda$ hypernuclei have been
studied with the antisymmetrized molecular dynamics (AMD)
model,\cite{Isaka11-1} in which the projection techniques and
generator coordinate method (GCM) have been implemented to take into
account the above deficiencies of static mean-field approaches. As
the gaussian overlap approximation of GCM, the collective Bohr
Hamiltonian with parameters determined by the self-consistent
mean-field calculations is much simple in numerical calculations,
and has achieved great success in description of the low-lying
states of normal nuclei.~\cite{Prochniak09} In view of these facts,
most recently, we have constructed a five-dimensional collective
Hamiltonian (5DCH) with the parameters derived from the SHF+BCS
calculations for the nuclear core $^{25}_\Lambda$Mg and calculated
the corresponding low-spin excitation spectra.~\cite{Yao11C} The
SGII force~\cite{Giai81} for the $NN$ effective interaction and the
No.1 set in Ref.~\cite{Yamamoto88} for the $\Lambda N$ interaction
were adopted. It has been shown that the $\Lambda$ hyperon stretches
the ground state band and reduces the $B(E2:2^+_1 \rightarrow
0^+_1)$ value by $\sim 9\%$, mainly by softening the potential
energy surface towards the spherical shape. Similar conclusion has
also been found in the AMD study for $^{25}_\Lambda$Mg quite
recently.\cite{Isaka11-2}

In this paper, we apply the same framework of Ref.~\cite{Yao11C} to
quantitatively evaluate the impurity effect of $\Lambda$ hyperon on
the collectivity of shape-coexistence nucleus $^{44}$S with several
different Skyrme forces for both the $NN$ and $\Lambda N$ effective
interactions as well as different pairing strengths for nucleons.

 \section{The Skyrme energy functional based collective Hamiltonian}
The framework of Skyrme energy functionals based collective
Hamiltonian for nuclear core of $\Lambda$ hypernuclei has been
explained in detail in Ref.~\cite{Yao11C}. Here, we just present an
outline. In this model, we start from the collective Hamiltonian
that describes the nuclear excitations of quadrupole vibrations, 3D
rotations, and their couplings~\cite{Libert99,Prochniak04,Li09}
\begin{equation}
\label{hamiltonian-quant} \hat{H} =
\hat{T}_{\textnormal{vib}}+\hat{T}_{\textnormal{rot}}
              +V_{\textnormal{coll}} \; ,
\end{equation}
where $V_{\textnormal{coll}}$ is the collective potential. The
vibrational kinetic energy reads,
\begin{eqnarray}
\hat{T}_{\textnormal{vib}}
 &=&-\frac{\hbar^2}{2\sqrt{wr}}
   \left\{\frac{1}{\beta^4}
   \left[\frac{\partial}{\partial\beta}\sqrt{\frac{r}{w}}\beta^4
   B_{\gamma\gamma} \frac{\partial}{\partial\beta}- \frac{\partial}{\partial\beta}\sqrt{\frac{r}{w}}\beta^3
   B_{\beta\gamma}\frac{\partial}{\partial\gamma}
   \right] \right.\nonumber\\
  &&  \left.+\frac{1}{\beta\sin{3\gamma}} \left[
   -\frac{\partial}{\partial\gamma} \sqrt{\frac{r}{w}}\sin{3\gamma}
      B_{\beta \gamma}\frac{\partial}{\partial\beta}
    +\frac{1}{\beta}\frac{\partial}{\partial\gamma} \sqrt{\frac{r}{w}}\sin{3\gamma}
      B_{\beta \beta}\frac{\partial}{\partial\gamma}
   \right]\right\},
 \end{eqnarray}
and the rotational kinetic energy,
\begin{equation}
\hat{T}_{\textnormal{\textnormal{\textnormal{rot}}}} =
\frac{1}{2}\sum_{\kappa=1}^3{\frac{\hat{J}^2_\kappa}{\mathcal{I}_\kappa}},
\end{equation}
with $\hat{J}_\kappa$ denoting the components of the angular
momentum in the body-fixed frame of a nucleus. The mass parameters
$B_{\beta\beta}$, $B_{\beta\gamma}$, $B_{\gamma\gamma}$, as well as
the moments of inertia $\mathcal{I}_\kappa$, depend on the
quadrupole deformation variables $\beta$ and $\gamma$,
\begin{equation}
 \label{MOI}
\mathcal{I}_\kappa = 4B_\kappa\beta^2\sin^2(\gamma-2\kappa\pi/3),
~~\kappa=1,2,3 \;.
\end{equation}
Two additional quantities that appear in the expression for the
vibrational energy, that is, $r=B_1B_2B_3$, and
$w=B_{\beta\beta}B_{\gamma\gamma}-B_{\beta\gamma}^2 $, determine the
volume element in the collective space. The dynamics of the
collective Hamiltonian is governed by seven collective quantities,
that is, the collective potential $V_{\rm coll}$, three mass
parameters $B_{\beta\beta}$, $B_{\beta\gamma}$, and
$B_{\gamma\gamma}$, and three moments of inertia
$\mathcal{I}_\kappa$, all of which are determined by the constrained
SHF+BCS calculations for the nuclei with and without $\Lambda$
hyperon. Recently, in Ref.~\cite{Win11}, the computer code {\tt
ev8}~\cite{Bonche05} of SHF+BCS approach has been extended for the
study of $\Lambda$ hypernuclei, which makes it feasible to carry out
the quantitative study of $\Lambda$ effect on nuclear collectivity.

In the SHF+BCS calculations for $\Lambda$ hypernucleus, the total
energy $E_{\rm tot}$ can be written as the integration of three
terms,
 \begin{equation}
 \label{HFE}
 E_{\rm tot}
 =\int d^3r [{\cal E}_N(\br) + {\cal T}_\Lambda(\br) + {\cal E}_{N\Lambda}(\br)],
 \end{equation}
where ${\cal E}_N(\br)$ is the standard nuclear part of Skyrme
energy functional.\cite{Bonche05,Vautherin72} ${\cal
T}_\Lambda(\br)=\dfrac{\hbar^2}{2m_\Lambda}\tau_\Lambda$ is the
kinetic energy density of $\Lambda$ hyperon. ${\cal
E}_{N\Lambda}(\br)$ is the interaction energy density between the
$\Lambda$ and nucleons given in terms of the $\Lambda$ and nucleon
densities,\cite{Rayet81}
\begin{eqnarray}
  {\cal E}_{N\Lambda}
  &=&
  t^\Lambda_0(1+\dfrac{1}{2}x^\Lambda_0)\rho_\Lambda\rho_N
  +\dfrac{1}{4}(t^\Lambda_1+t^\Lambda_2)(\tau_\Lambda\rho_N+\tau_N\rho_\Lambda)\nonumber\\
  &&+\dfrac{1}{8}(3t^\Lambda_1-t^\Lambda_2)(\nabla\rho_N\cdot\nabla\rho_\Lambda)
  +\dfrac{1}{4}t^\Lambda_3\rho_\Lambda(\rho^2_N+2\rho_n\rho_p)\nonumber\\
  &&+\dfrac{1}{2}W^\Lambda_0(\nabla\rho_N\cdot \bJ_\Lambda+\nabla\rho_\Lambda\cdot \bJ_N)
  \tau_N\rho_\Lambda.
\end{eqnarray}
Here, $\rho_\Lambda, \tau_\Lambda$ and $\bJ_\Lambda$ are
respectively the particle density, the kinetic energy density, and
the spin density of the $\Lambda$ hyperon. These quantities are
given in terms of the single-particle wave-function of $\Lambda$ and
occupation probabilities.\cite{Vautherin72} $t^\Lambda_0,
t^\Lambda_1, t^\Lambda_2, t^\Lambda_3$, and $W^\Lambda_0$ are the
Skyrme parameters for the $\Lambda N$ interaction.

The pairing correlation between the nucleons is taken into account
in the BCS approximation. The density-dependent $\delta$-force is
adopted in the $pp$ channel,
\begin{equation}
 V(\br_1, \br_2)=-g\dfrac{1-\hat P^\sigma}{2}
 \left[1-\dfrac{\rho(\br_1)}{\rho_0}\right]
 \delta(\br_1-\br_2),
\end{equation}
where $\hat P^\sigma$ is the spin-exchange operator, and
$\rho_0=0.16$ fm$^{-3}$.

\section{Results and discussion}

\subsection{Different Skyrme forces for particle-hole channel}

\begin{figure}[]
\centerline{\psfig{file=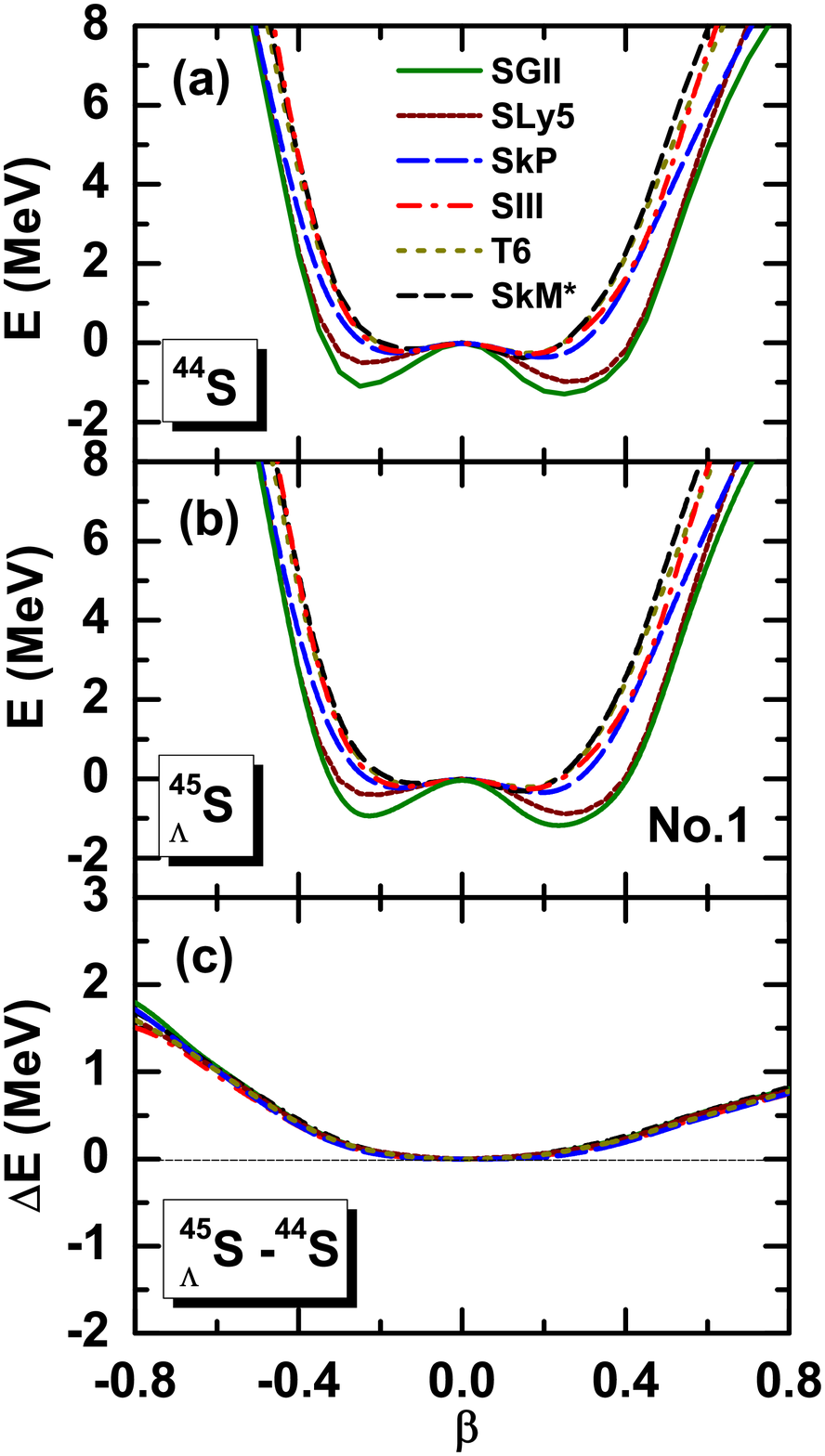,width=4.5cm}} %\vspace*{8pt}
\caption{(color online) Potential energy curves (PECs) obtained by
the SHF+BCS calculations with several different Skyrme forces but
the same pairing strength ($g=1000$ MeV fm$^3$) for (a) $^{44}$S and
(b) $^{45}_\Lambda$S. The difference between the PECs of
$^{45}_\Lambda$S  and $^{44}$S is normalized to the spherical shape
and shown in (c). } \label{fig1}
\end{figure}

Figure~\ref{fig1} shows the potential energy curves (PECs) from the
SHF+BCS calculations with several Skyrme forces for $^{44}$S and
$^{45}_\Lambda$S, where the No.1 set of $\Lambda N$ effective
interactions in Ref.~\cite{Yamamoto88} and pairing strength $g=1000$
MeV fm$^3$ are adopted. It is shown that the PECs of $^{44}$S by the
SGII~\cite{Giai81} and SLy5~\cite{Chabanat98} forces have two
obvious minima in both oblate and prolate sides. The PECs by other
Skyrme forces are very similar and soft along $\beta$ direction
around the spherical shape. Similar situation is also found in
$^{45}_\Lambda$S. Even though these Skyrme forces predict somewhat
different PECs for $^{44}$S, the resultant polarization effects of
$\Lambda$ by these forces are quite similar, as shown in the panel
(c) of Fig.~\ref{fig1}. Moreover, the behavior of the difference
between the PECs of $^{44}$S and $^{45}_\Lambda$S indicates that the
$\Lambda$ has the effect of driving the nucleus to be of small
deformation, i.e., reducing the nuclear collectivity.

\begin{figure}[]
\centerline{\psfig{file=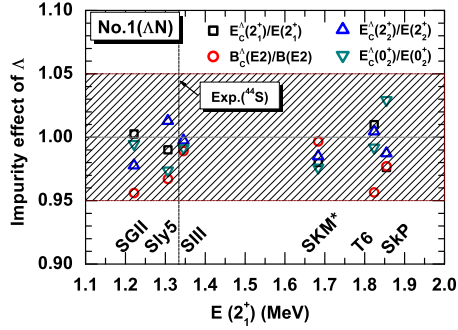,width=7cm}} %\vspace*{8pt}
\caption{(color online) The ratios of spectroscopic observables for
nuclear core of $^{45}_\Lambda$S to those for $^{44}$S from the 5DCH
calculations with parameters determined by the SHF+BCS calculations
with several Skyrme forces as functions of excitation energy of
$2^+_1$ state in corresponding normal nuclei. } \label{fig2}
\end{figure}

In Fig.~\ref{fig2}, we plot the ratios of spectroscopic observables
for nuclear core of $^{45}_\Lambda$S to those for $^{44}$S from the
5DCH calculations with parameters determined by the SHF+BCS
calculations with several different Skyrme forces. The experimental
excitation energy of $2^+_1$ state ($E(2^+_1)$) in $^{44}$S is
indicated with a vertical dotted line. We note that the $E(2^+_1)$
of $^{44}$S by the SGII, SLy5 and SIII~\cite{Beiner75} forces are
much smaller than those by SKM*, T6 and SkP forces, which can be
understood from the PECs in Fig.~\ref{fig1}, except the case of SIII
force. Namely, the collective wavefunctions of $2^+_1$ state by the
calculations of SGII and SLy5 forces are more concentrated on the
obvious deformed minima of the PECs with larger moments of inertia.
On the other hand, it shows clearly that the $B(E2:2^+_1\rightarrow
0^+_1)$ is reduced by the $\Lambda$ hyperon. In particular, these
different Skyrme forces give similar results of $\Lambda$ impurity
effect in $^{44}$S, i.e., the $\Lambda$ impurity effect on
collectivity of $^{44}$S is within $5\%$.

\begin{figure}[htp]
\centerline{\psfig{file=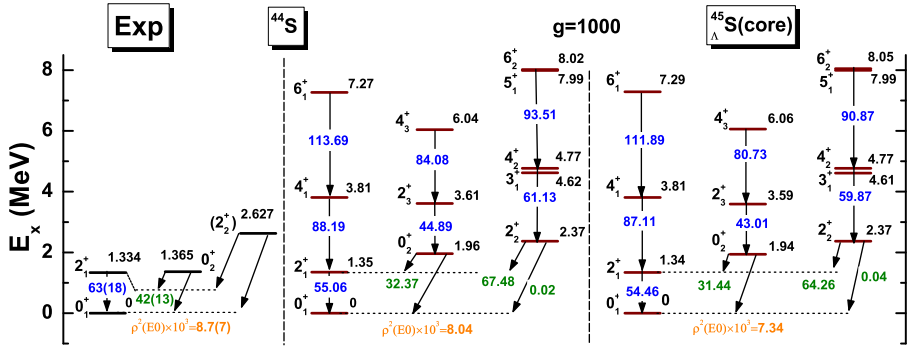,width=12cm}} \caption{(color
online) The low-spin spectra of the ground state band for the
$^{44}$S and the nuclear core of $^{45}_\Lambda$S obtained by the
5DCH with the parameters determined by the SHF+BCS calculations
using the SIII ($NN$) and No.1 ($\Lambda N$) Skyrme forces. The
$B(E2)$ values are in units of e$^2$ fm$^4$. The spectrum of
$^{44}$Mg is compared with the corresponding available experimental
data. } \label{fig3}
\end{figure}

Figure~\ref{fig3} displays the low-spin spectra for the $^{44}$S and
the nuclear core of $^{45}_\Lambda$S obtained by the 5DCH
calculations with the parameters determined by the calculations of
SIII~\cite{Beiner75} ($NN$) and No.1~\cite{Yamamoto88} ($\Lambda N$)
Skyrme forces. The experimental data for $^{44}$S are taken from
Refs.~\cite{Glasmacher97,Force10}. It is shown that the spectrum of
$^{44}$S is reproduced very well and the effect of $\Lambda$ hyperon
on the low-spin states in $^{44}$S is generally small.

\subsection{Different pairing strengths for particle-particle channel}

\begin{figure}[]
\centerline{\psfig{file=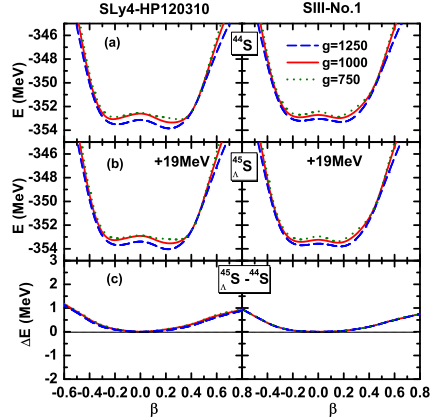,width=6cm}} %\vspace*{8pt}
\caption{(color online) The potential energy curves obtained by the
Skyrme HF+BCS calculations with both the SIII and SLy4 EDFs, and
different pairing strengths for (a) $^{44}$S and (b)
$^{45}_\Lambda$S. The difference between the PECs of
$^{45}_\Lambda$S and $^{44}$S, normalized to the spherical shape, is
shown in (c). } \label{fig4}
\end{figure}
\begin{figure}[]
\centerline{\psfig{file=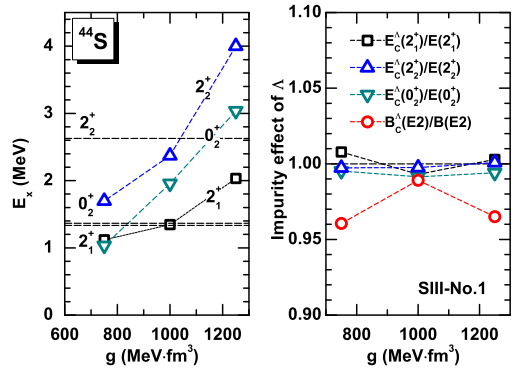,width=7cm}} %\vspace*{8pt}
\caption{(color online) (Left panel) The excitation energies of
$2^+_1, 0^+_2$ and $2^+_2$ states in $^{44}$S and (right panel) the
ratios of spectroscopic observables for nuclear core of
$^{45}_\Lambda$S to those for $^{44}$S as functions of pairing
strength $g$ for nucleons. The experimental data for $^{44}$S are
indicated with dashed lines in the left panel.} \label{fig5}
\end{figure}

Figure~\ref{fig4} shows the PECs obtained by the SHF+BCS
calculations with both the SIII and SLy4~\cite{Chabanat98} forces,
and with different pairing strengths for $^{44}$S and
$^{45}_\Lambda$S. The No.1 (adjusted based on the SIII) of
Ref.~\cite{Yamamoto88} and the ``HP120310" (adjusted based on the
SLy4) of Ref.~\cite{Guleria11} are adopted for $\Lambda N$ effective
interactions respectively. It is shown that the pairing strengths
for nucleons do have influence on the topology of PECs for $^{44}$S.
However, it has a negligible effect on the polarization effect of
$\Lambda$ hyperon. Moreover, it is shown again that the polarization
effect of $\Lambda$ hyperon in $^{44}$S is similar for the SIII and
SLy4 forces.

Subsequently, we perform the further 5DCH calculations using only
the SIII force, but different pairing strengths for nucleons.
Figure~\ref{fig5} displays the corresponding spectroscopic
properties of low-spin states for $^{44}$S and nuclear core of
$^{45}_\Lambda$S as functions of pairing strength $g$ for nucleons.
It shows clearly and quantitatively that the excitation energies of
$2^+_1, 0^+_2$ and $2^+_2$ states in $^{44}$S increase monotonically
and dramatically with the pairing strengths, which can be understood
that the strong pairing generally drives the nucleus to be more
spherical. Similar phenomenon has also been shown in the covariant
EDF based 5DCH calculations.\cite{Li11} However, the impurity effect
of $\Lambda$ on nuclear collectivity is not much sensitive to the
pairing strength for nucleons, which varies within $5\%$ for
different pairing strengths.

 \section{Summary and outlook}
The non-relativistic Skyrme EDF based 5DCH, that takes into account
dynamical correlations related to the restoration of broken
symmetries and fluctuations of quadrupole collective variables, has
been applied to quantitatively study the impurity effect of
$\Lambda$ hyperon on collective excitation of shape-coexistence
nucleus $^{44}$S. Several Skyrme forces for both the $NN$ and
$\Lambda N$ effective interactions have been used. The influence of
pairing strengths on the polarization effect of $\Lambda$ hyperon
has been also examined. It has been found that although these Skyrme
forces give somewhat different low-lying spectra for $^{44}$S, they
give similar and generally small size of $\Lambda$ polarization
effect (within $5\%$) on the spectroscopic observables. Moreover,
the pairing strength between nucleons has significant influence on
the nuclear collectivity, but negligible effect on the polarization
effect of $\Lambda$ hyperon.

With the Skyrme EDF based 5DCH model, the systematic study of
$\Lambda$ impurity effect on collectivity of atomic nuclei in
different mass region is in progress.\cite{Mei11} Moreover, as
pointed out in Refs.~\cite{Schulze10,Win08}, the polarization effect
of $\Lambda$ hyperon might be stronger in the calculations with
covariant EDFs. The quantitative evaluation of $\Lambda$ effect on
nuclear collectivity will provide a good way to constrain $\Lambda
N$ effective interaction in covariant EDF. Furthermore, to calculate
the low-spin spectra for the whole single $\Lambda$ hypernucleus,
one has to extend the current EDF based 3DAMP+GCM~\cite{Yao09,Yao10}
or 5DCH models for odd-mass/-odd nuclei.

\section*{Acknowledgements}

JMY thanks H. F. L\"{u}, J. Meng, C. Y. Song, M. Thi Win, and Y.
Zhang for their collaborations on the studies of hypernuclei, thanks
P.-H. Heenen and K. Washiyama for discussions about the
non-relativistic mean-field calculations and S. G. Zhou for
discussions about the relativistic results and acknowledges the
postdoctoral fellowship from the F.R.S.-FNRS (Belgium). This work
was partly supported by the NSFC under Grants No. 11105111, No.
11105110 and No. 10947013, the Fundamental Research Funds for the
Central Universities (XDJK2010B007 and XDJK2011B002), the Southwest
University Initial Research Foundation Grant to Doctor (SWU109011
and SWU110039) and the Japanese Ministry of Education, Culture,
Sports, Science and Technology by Grant-in-Aid for Scientific
Research under Program No. 22540262.

\end{document}